\documentclass[dvipsnames,sigplan]{acmart}
\acmConference[TyDe '20]{ACM SIGPLAN Workshop on Type-Driven Development}{August
23, 2020}{Remote}
\acmYear{2020}
\acmISBN{} 
\acmDOI{} 
\startPage{1}

\setcopyright{none}
\usepackage{booktabs}   
\usepackage{subcaption} 
\usepackage{algorithm2e}
\usepackage{lipsum}
\usepackage{listings}
\usepackage{xcolor}
\usepackage{float}

\definecolor{mygreen}{rgb}{0,0.6,0}
\definecolor{mygray}{rgb}{0.5,0.5,0.5}
\definecolor{mymauve}{rgb}{0.58,0,0.82}

\lstdefinestyle{holec}{language=C,
  backgroundcolor=\color{white},   
  basicstyle=\ttfamily\small,        
  breakatwhitespace=false,         
  breaklines=true,                 
  captionpos=b,                    
  commentstyle=\color{mygreen},    
  extendedchars=true,              
  keepspaces=true,                 
  numbers=left,                    
  numbersep=5pt,                   
  numberstyle=\tiny\color{mygray}, 
  rulecolor=\color{black},         
  showspaces=false,                
  showstringspaces=false,          
  showtabs=false,                  
  stringstyle=\color{mymauve},     
  tabsize=2,	                   
  keywordstyle=\color{blue},       
  keywordstyle = [2]{\bfseries},
  keywordstyle = [3]{\bfseries\color{blue}},
  otherkeywords={??,?T,?U,?V},
  morekeywords = [2]{??},
  morekeywords = [3]{?T,?U,?V}
}

\lstdefinestyle{llvm}{language=llvm,
  backgroundcolor=\color{white},   
  basicstyle=\ttfamily\small,        
  breakatwhitespace=true,         
  breaklines=true,                 
  postbreak=\mbox{\textcolor{red}{$\hookrightarrow$}\space},
  captionpos=b,                    
  commentstyle=\color{mygreen},    
  extendedchars=true,              
  keepspaces=true,                 
  numbers=left,                    
  numbersep=5pt,                   
  numberstyle=\tiny\color{mygray}, 
  rulecolor=\color{black},         
  showspaces=false,                
  showstringspaces=false,          
  showtabs=false,                  
  stringstyle=\color{mymauve},     
  tabsize=2,	                   
  keywordstyle=\color{blue},       
  keywordstyle = [2]{\color{BrickRed}},
  otherkeywords = {type},
  morekeywords = [2]{type}
}
\newcommand{\llvmin}[1]{\lstinline[style=llvm]{#1}}

\newcommand{\rauw}{\textsc{rauw}}
\newcommand{\rauwnt}{\textsc{rauw-nt}}

\begin{document}

\title[]{Retrofitting Symbolic Holes to LLVM IR}
\subtitle{(Extended Abstract)}

\author{Bruce Collie}
\affiliation{
  \institution{University of Edinburgh}
}
\email{bruce.collie@ed.ac.uk}

\author{Michael O'Boyle}
\affiliation{
  \institution{University of Edinburgh}
}
\email{mob@inf.ed.ac.uk}

\begin{abstract}

Symbolic holes are one of the fundamental building blocks of solver-aided and
interactive programming. Unknown values can be soundly integrated into programs,
and automated tools such as SAT solvers can be used to prove properties of
programs containing them.  However, supporting symbolic holes in a programming
language is challenging; specifying interactions of holes with the type system
and execution semantics requires careful design.

This paper motivates and introduces the implementation of symbolic holes with
unknown type to LLVM IR, a strongly-typed compiler intermediate language. We
describe how such holes can be implemented safely by abstracting unsound and
type-unsafe details behind a new primitive IR manipulation. Our implementation
co-operates well with existing features such as type and dependency checking.
Finally, we highlight potentially fruitful areas for investigation using our
implementation.

\end{abstract}

\maketitle

\section{Context}

Our work in this paper is motivated by efforts to extend a program synthesizer
we implemented in previous work \cite{Collie2019,Collie2019a}. This synthesizer
produces programs in LLVM intermediate representation (IR) \cite{Lattner2002};
we do this to allow for synthesized programs to be inserted into existing
applications, and to permit translation to a searchable representation
\cite{Ginsbach2018a}.

In \cite{Collie2019a} we synthesize programs by first generating control-flow
code based on a library of partial components, then stochastically inserting
instructions to each generated basic block. This approach did not scale as the
complexity of synthesized programs increased; there were too many possible
locations that instructions could be added to, and we had no general way to
express constraints on what instructions should be sampled.

The solution we arrived at was for components to contain \emph{symbolic holes}.
Instead of selecting both location and value for instructions, the synthesiser
now only has to select a value. Our design constraints can therefore be
summarised: we require a way to add symbolic holes to LLVM IR programs, and an
interface by which values can be assigned to these holes. Some of our components
are generic, so holes with no explicit type should be supported.  Finally, the
implementation should remain as compatible with existing LLVM tools for program
manipulation (i.e.\ programs with holes should still be valid IR).

\section{Encoding Holes}

Our encoding of holes uses ``uninterpreted'' functions with no definition to
represent symbolic holes; one such function declaration is generated for each
symbolic hole. For example, a hole of type \llvmin{i32} is encoded as:
\begin{lstlisting}[language=llvm, style=llvm]
declare i32 @hole0()
%0 = call i32 @hole0()
\end{lstlisting}

When manipulating this program, the value \llvmin{\%0} can be used anywhere a
concrete value of type \llvmin{i32} could be:
\begin{lstlisting}[language=llvm, style=llvm]
%1 = add i32 %0, i32 1
\end{lstlisting}

For holes where the type is not known ahead of time, we create a special hole
type:
\begin{lstlisting}[language=llvm, style=llvm]
%hole.t = type {}
declare %hole.t @hole1()
%2 = call %hole.t @hole1()
\end{lstlisting}

If a hole is known to depend on other values (hole or not), we encode this using
function parameters. For example, if we know that \llvmin{\%3} should depend on
\llvmin{\%1} and \llvmin{\%2}, but not specifically how it should be computed:
\begin{lstlisting}[language=llvm, style=llvm]
declare %hole.t @hole2(i32, %hole.t)
%3 = call %hole.t @hole2(i32 %1, %hole.t %2)
\end{lstlisting}

To allow for hole values of unknown type to appear in concrete operations (e.g.\
computing the sum of two hole values), we use a similar encoding. Actual LLVM
opcodes such as \llvmin{add} do not support custom types like \llvmin{\%hole.t}:
\begin{lstlisting}[language=llvm, style=llvm]
%4 = call %hole.t @add(%hole.t %2,    %hole.t %3)
\end{lstlisting}

These functions are replaced by concrete opcodes once the type of both operands
is known.

This encoding of holes and operations allows us to remain safely within the LLVM
type system, and to take full advantage of safety mechanisms such as use-def
checking. Programs using our encoding are valid LLVM and can be manipulated as
such, but cannot yet be linked and executed.

\section{Solver Interface}

The programs we encode with symbolic holes are not yet complete. A concrete
value must be assigned to each hole in order to produce an executable program.
These assignments are the role of a domain-specific client (e.g.\ a solver or
synthesizer). Our encoding of holes is independent of the decision procedure
that assigns values to them.

LLVM makes frequent internal use of the ``replace all uses with'' (\rauw{})
primitive. \rauw{} replaces a value in a program with another of the same type;
because LLVM IR is in SSA form this replacement is well-defined.  If the type of
all holes are known, then clients would simply be able to use \rauw{} to
substitute each hole for an appropriate value. However, we allow the type of
holes to be unknown and so require a more powerful abstraction.

We introduce a new IR manipulation primitive: \rauwnt{} (New Type) that
abstracts a set of unsound transformations that effectively change the type of
values.

\paragraph{Implementing \rauwnt{}}

When the original and replacement values have the same type, \rauwnt{} behaves
identically to \rauw{}. When they do not, \rauwnt{} behaves as follows.

First, it checks that the original value has type \llvmin{\%hole.t}. Arbitrary
changes of type are not supported; fixing a value for a hole with unknown type
is the only supported case. Then, any use (e.g.\ a hole depending on it, or an
operation on holes such as \llvmin{@add}) of the original value is redeclared
with the relevant parameter type changed. The call site is then replaced with
one where the new value is passed. Finally, the original holes are deleted and
the new values renamed where appropriate.

In the code below, consider assigning the constant \llvmin{i32 5} to the hole
\llvmin{\%0}:
\begin{lstlisting}[language=llvm, style=llvm]
declare %hole.t @hole()
declare i32 @hole1(%hole.t)
%0 = call %hole.t @hole()
%1 = call i32 @hole1(%hole.t %0)
\end{lstlisting}

The code produced by \rauwnt{} will (after renaming) be:
\begin{lstlisting}[language=llvm, style=llvm]
declare i32 @hole1(i32)
%1 = call i32 @hole1(i32 5)
\end{lstlisting}

The original hole no longer exists, and its uses appear to have changed type to
\llvmin{i32}.

\paragraph{Backpropagating Types}

If a value is assigned to a hole that is used by an operation, then type
information may flow backwards as well as forwards. For example, we know that
both operands of an add operation must be the same type, and the result also has
that type. We therefore replace operands and operations with typed ones when
partial information becomes available about them. It is also possible for
\rauwnt{} to fail at this point if the inferred types are incompatible.

Implementing \rauwnt{} requires manipulating LLVM IR in a way not originally
intended by its designers, and is fiddly to implement correctly. By abstracting
it, clients can make high-level decisions on the values they assign without
worrying about low-level IR manipulation.

\section{Research Directions}

As described, our method for embedding holes in LLVM IR programs is only the
first implementation step towards full solver-aided programming integrated with
the compiler. Some promising next steps towards this are:

\paragraph{Synthesis}

We are already using our implementation in the next version of our synthesizer;
in a paper currently under review we are able to synthesize more complex
functions than in previous work \cite{Collie2019a}. Being able to express
constraints and partial type information in synthesis components independently
of each other has proved to be a useful feature.

\paragraph{Solver Integration}

We hope to investigate further integration with solver-aided techniques beyond
whole-program synthesis. Where our synthesis procedure uses whole-program
behaviour to determine correctness, solver-aided techniques generally use local
measures such as assertions; these can be added independently of our hole
encoding. One use of these assertions is to implement \emph{angelic execution},
one of the principal techniques highlighted in \cite{Torlak2013}

Another example is superoptimization over bit-vector programs; these problems
can be stated easily within the LLVM type system. Expressing a length-$N$
superoptimization constraint can be achieved simply by using $N$ suitably-typed
symbolic holes.

\paragraph{Source Language Support}

Another avenue for future work is the development of libraries in languages that
target LLVM IR (e.g.\ C, C++ or Fortran) that expose symbolic hole primitives to
programmers. Doing so would allow for programs written in these languages to
take advantage of solver-aided programming without requiring specialised
knowledge of the compiler.

\paragraph{Tool Support}

Languages with first-class hole support such as Hazel \cite{Omar2019} emphasize
the need for supporting tooling and programming environments. Extending this
kind of support to LLVM IR programs would enable interactive programming for a
wide class of existing applications.

\paragraph{Theory}

This paper focuses on the implementation of a system for manipulating LLVM IR
programs in the presence of values with unknown type; the implementation is
driven by the constraints of the existing language tooling and manipulation
utilities. Existing work \cite{Zhao2012a} deals with formalising LLVM IR;
grounding our system formally in similar terms alongside gradual typing and
hole-enabled programming is interesting future work.

\clearpage

\bibliography{references.bib}

\end{document}